\documentclass[aps,prd,superscriptaddress,eqsecnum,nofootinbib,showkeys,twocolumn,showpacs,preprintnumbers]{revtex4}
\usepackage{graphicx}
\usepackage{euscript,amssymb}
\usepackage{amsfonts}
\usepackage{amsmath}
\usepackage{amssymb}
\usepackage{graphicx}
\usepackage{euscript}
\usepackage{color}
\begin{document}

\title{Cosmological Imprints of a Generalized Chaplygin Gas Model for the Early Universe}

\author{Mariam Bouhmadi-L\'{o}pez}\email{mariam.bouhmadi@ist.utl.pt}
\affiliation{Centro Multidisciplinar de Astrof\'{\i}sica - CENTRA, Departamento de F\'{\i}sica, Instituto Superior T\'ecnico, Av. Rovisco Pais 1,1049-001 Lisboa, Portugal}
\author{Pisin Chen}\email{chen@slac.stanford.edu}
\affiliation{Department of Physics \& Graduate Institute of Astrophysics, National Taiwan University, Taipei 10617, Taiwan, R.O.C.}
\affiliation{Leung Center for Cosmology and Particle Astrophysics,
National Taiwan University, Taipei 10617, Taiwan, R.O.C.}
\affiliation{Kavli Institute for Particle Astrophysics and Cosmology,
SLAC National Accelerator Laboratory, Menlo Park, CA 94025, U.S.A.}
\author{Yen-Wei Liu}\email{f97222009@ntu.edu.tw}
\affiliation{Department of Physics \& Graduate Institute of Astrophysics, National Taiwan University, Taipei 10617, Taiwan, R.O.C.}
\affiliation{Leung Center for Cosmology and Particle Astrophysics,
National Taiwan University, Taipei 10617, Taiwan, R.O.C.}
\begin{abstract}
We propose a phenomenological model for the early universe where there is a
smooth transition between an early ``quintessence'' phase and a radiation dominated era.
The matter content is modelled  by an appropriately  modified Chaplygin gas  for the early universe. We constrain the model observationally by mapping the primordial power spectrum of the scalar perturbations to the latest data of WMAP7. We compute as well the spectrum of the primordial gravitational waves as would be measured today. We show that the high frequencies region of
the spectrum depends on the free parameter of the model and most importantly this region of the spectrum can be within the reach of future gravitational waves detectors.

\end{abstract}

%\pacs{98.80.-k,98.80.Es,11.10.-z}

\date{\today}
\maketitle

%\begin{document}

\section{Introduction}

While our knowledge about the universe has improved over the last decades with the advent of new observational data, there are several dark sides of the universe that have not been so far described from a fundamental point of view: what caused the initial inflationary era of the universe? what is the origin of dark matter? what is the fundamental cause of the current acceleration of the universe? Even though nowadays none of the previous issues have  a satisfactory answer, a parallel approach, that can shed some light on the dark sides of the universe, is a phenomenological one or a model building strategy. A good example in this regard is the inflationary paradigm \cite{inflation}, where an inflaton field (or several scalar fields) induce the initial acceleration of the universe. Such a field would leave tracks on the  evolution of the universe for example through the primordial power spectrum of the scalar perturbations \cite{Langlois:2010xc,Bassett:2005xm,Lidsey:1995np}, extremely useful to constrain the inflaton field. It is precisely such an approach that we will follow in this paper.

The main goal of this paper is to obtain a phenomenologically consistent model for the early universe (inflationary and radiation dominated epochs) by properly modifying  the generalised Chaplygin gas (GCG) \cite{chaplygin}.
A first attempt in this direction has been recently carried out in \cite{BouhmadiLopez:2009hv} (see also \cite{Bertolami:2006zg}) where a new scenario for the early universe was proposed. Such a scenario provides a  smooth transition between an early de Sitter-like phase and a subsequent radiation dominated era. In that model \cite{BouhmadiLopez:2009hv}, the matter content was given by a type of generalized Chaplygin gas for the early universe, with an underlying scalar field description. Here, we give a more \textit{realistic} model where the early inflationary phase of the universe is more general than a de Sitter-like universe. We will show how the spectrum of the present model is not as red as the one presented in \cite{BouhmadiLopez:2009hv} and therefore more consistent with observations.

More precisely, rather than having a de Sitter-like phase in the past, we will consider an early  ``quintessence'' inflationary phase. This phase will be connected to a radiation dominated phase at later time. The model can be described through a scalar field or a Chaplygin gas inspired model. We will then analyze the possible imprints of such a gas in the primordial power spectrum of scalar perturbations and the power spectrum of the stochastic background of gravitational waves.

The paper is organized as follows. In section II, we will present the model which is based on a properly modified Chaplygin gas that interpolates between a quintessence inflationary era and a radiation dominated period. In section III, we will show how this kind of gas can be modeled by a minimally coupled scalar field. In section IV, we constrain our model observationally and obtain the full spectrum of scalar perturbations numerically. In section V, we obtain the spectrum of gravitational waves using the method of Bogoliubov coefficient. Finally, in section VI we summarize our results and conclude.

\section{The model building}

We generalize the model presented in \cite{BouhmadiLopez:2009hv} by considering an inflationary period corresponding to a ``quintessence'' like behavior (described by a power law expansion) and  followed by a radiation dominated epoch. The matter content of the universe can then be modeled \textit{\`a la} Chaplygin gas as
\begin{equation}
\rho=\left(\frac{A}{a^{1+\beta}}+\frac{B}{a^{4(1+\alpha)}}\right)^{1/(1+\alpha)},
\label{rho}
\end{equation}
where $A,B,\alpha,\beta$ are constants being $A, B$ positive.

The matter content is not interacting with any other fluid and therefore its energy density is conserved:
\begin{equation}
\dot\rho+3H(\rho+p)=0,
\label{conservation}
\end{equation}
where the dot stands for derivative with respect to the cosmic time and $p$ is the pressure of the fluid. If we substitute the energy density (\ref{rho}) in the conservation equation (\ref{conservation}), we obtain
\begin{equation}
p=\frac13\rho+\frac{1+\beta-4(1+\alpha)}{3(1+\alpha)}
\left(\rho-\frac{B}{a^{4(1+\alpha)}}\rho^{-\alpha}\right).
\label{p}
\end{equation}
The above equation of state can be rewritten as
\begin{equation}
p=\frac13\rho+\frac{A}{3(1+\alpha)}\frac{1+\beta-4(1+\alpha)}{a^{1+\beta}} \rho^{-\alpha}.
\label{p2}
\end{equation}
This equation shows clearly that we recover the model discussed by one of us in \cite{BouhmadiLopez:2009hv} for $\beta\rightarrow - 1$. We would like to highlight that the equations of state (\ref{p})-(\ref{p2}) have been previously analyzed in \cite{Chimento:2009sh} to study a possible interplay between dark matter and dark energy and therefore in a scenario completely different to the one we study here\footnote{We thank Luis Chimento for pointing out this to us.}.

As the inflationary period takes place before the radiation dominated one, we deduce that the inequality
\begin{equation}
  \frac{B}{a^{4(1+\alpha)}}\ll\frac{A}{a^{1+\beta}}
\label{inequality1}
\end{equation}
must hold at early time; i.e. at small scale factors. This will be the case as long as
\begin{equation}
0<1+\beta-4(1+\alpha).
\label{inequality2}
\end{equation}
The inequality (\ref{inequality1}) implies $a\ll a_{\rm{l}}$, where
\begin{equation}
a_{\rm{l}}= \left(\frac{A}{B}\right)^{1/[1+\beta-4(1+\alpha)]}.
\end{equation}
The scale factor $a_{\rm{l}}$ represents the ``border line'' between two different regimes. More precisely
\begin{eqnarray}
  \rho&\simeq& \frac{A^{1/(1+\alpha)}}{a^{(1+\beta)/(1+\alpha)}} \quad \,\,a\ll a_{\rm{l}}, \label{rhoinf}\\
  \rho&\simeq& \frac{B^{1/(1+\alpha)}}{a^4} \quad a\gg a_{\rm{l}}.\label{rhorad}
\end{eqnarray}

There are two more conditions that should be imposed on $\alpha$ and $\beta$ to have a model that interpolates between an early inflationary phase of the type of quintessence and a radiation dominated phase at later time:
(i) the energy density (\ref{rhoinf}) must induce a period of inflation and (ii) such a period of inflation should not induce a super-inflationary expansion; i.e. $0<\dot H$, and therefore a super-accelerating phase of the universe where the energy density (\ref{rhoinf}) corresponds to phantom matter, with an energy density that grows as the universe expands. By combining these two ansatz with the inequality (\ref{inequality2}), we can easily deduce that the set of allowed values of $\alpha$ and $\beta$ satisfy\footnote{Having a negative value $\alpha$ is not that surprising (see for example \cite{BouhmadiLopez:2004me}). Indeed, it has been shown in \cite{Bertolami:2004ic} that type Ia Supernovae allow for negative $\alpha$ under the standard generalized Chaplygin gas.}
\begin{eqnarray}
&& 1+\alpha<0, \nonumber \\
&& 1+\beta<0, \nonumber \\
&& 2(1+\alpha) <1+\beta.
\label{conditions}
\end{eqnarray}

A perfect fluid with an equation of state (\ref{p}) or (\ref{p2}) describe an early inflationary period as long as the conditions (\ref{conditions}) are satisfied. The universe will exit the inflationary epoch when the scale factor reaches the value
\begin{equation}
a_{\star}=\left[\left(\frac{2(1+\alpha)-(1+\beta)}{1+\alpha}\right)\frac{A}{2B}\right]^{1/[1+\beta-4(1+\alpha)]}.
\end{equation}
The last condition can be obtained by imposing the condition $\rho+3p=0$. Again this is in agreement with the results of \cite{BouhmadiLopez:2009hv} for $\beta\rightarrow - 1$.

Before concluding this section, let us analyze more closely what sort of inflation does the universe undergo in its initial stages. The Friedmann equation with the matter content (\ref{rho}) is too difficult to be integrated analytically so we use an approximation where the energy density can be written as in Eq.~(\ref{rhoinf}). Then the scale factor can be approximated as
\begin{equation}
a(t)=\left[\frac{1+\beta}{2(1+\alpha)}A^{1/[2(1+\alpha)]}\sqrt{\frac{\kappa^2}{3}}t\right]^{2(1+\alpha)/(1+\beta)},
\label{powerlaw}
\end{equation}
where $t$ is the cosmic time and $\kappa^2=8\pi \rm{G}$ where $\rm{G}$ is the gravitational constant. As we can see from the previous equation the inflationary era of the universe is given by a power law expansion.

In sections IV and V, we will see that the entire evolution of the universe can be described by the appropriately modified Chaplygin gas (\ref{p2})  and a subsequent $\Lambda$CDM expansion, that is

\begin{eqnarray} \rho = \left\{ \begin{array}{lll}
          &\left(\frac{A}{a^{1+\beta}}+\frac{B}{a^{4(1+\alpha)}}\right)^{1/(1+\alpha)}, 
          &\,\,\mbox{early-time} \\
          & \,\, & \\
          &\rho_{r0}(\frac{a_0}{a})^4+\rho_{m0}(\frac{a_0}{a})^3+\rho_\Lambda,  & \,\,\mbox{late-time }\end{array} \right.  \label{energydensities}\end{eqnarray}
where $a_0$ is the current scale factor, $\rho_{r0}$, $\rho_{m0}$ and $\rho_\Lambda$ are the current energy densities corresponding to radiation, matter (cold dark matter and  baryonic matter)  and  dark energy (modelled through a cosmological constant), respectively. On the other hand, at the radiation dominated epoch the energy densities (\ref{energydensities}) are equal, consequently
\begin{equation}
B=\left(\rho_{r0}a_0^4\right)^{1+\alpha}\label{Bdef}.
\end{equation}
The parameter $A$ is related to the scale of inflation (see Eqs.~(\ref{rhoinf}) and (\ref{Vapowerlow})).

\begin{figure}[t]
  \includegraphics[width=8cm]{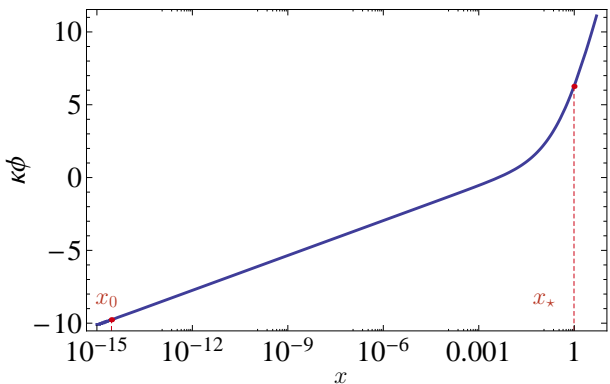}
	\caption{The blue curve corresponds to the scalar field, $\phi$, against $x$, where $x$ corresponds to a power of the scale factor $a$, $x=(B/A)a^q$, with $q=1+\beta-4(1+\alpha)$. The values $x_0$ and $x_\star$ correspond to the moments when the pivot scale $k_0=0.002\, \textrm{Mpc}^{-1}$ exists the horizon and the inflation ends, respectively.}
	\label{phixa}
\end{figure}

\section{An underlying scalar field model}

\begin{figure}[t]
  \includegraphics[width=8cm]{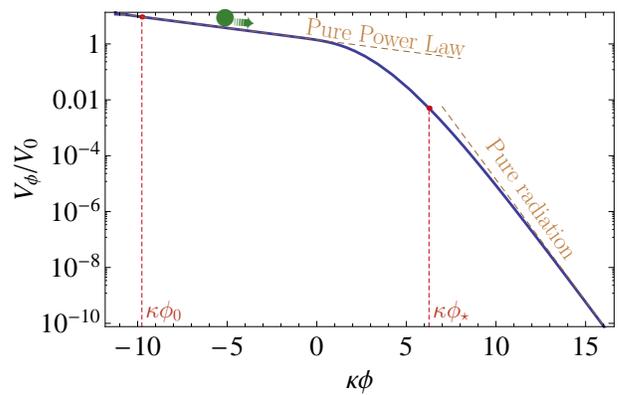}
	\caption{The blue curve corresponds to the scalar field potential $V(\phi)$ against $\phi$ (see Eqs.~(\ref{phia})-(\ref{Va})), where $V_0=A^{1/(1+\alpha)}\left(A/B\right)^{-(1+\beta)/[q(1+\alpha)]}$. The values $\phi_0$ and $\phi_\star$ correspond to the moments when the pivot scale $k_0=0.002\, \textrm{Mpc}^{-1}$ exists the horizon and the end of inflation, respectively.}
%The green curve stands for the power law potential given in Eq.~(\ref{Vapowerlow}). As can be seen the power law potential is a very good approximation of the model at very early time. The vertical left hand side line corresponds to the moment when the pivot mode $k_0$ exits the horizon while the right hand side vertical line corresponds to the end of the inflationary era.}
	\label{V1}
\end{figure}

The inflationary dynamics of the model presented in the previous section can be  described through a minimally coupled scalar field, $\phi$, with a potential, $V(\phi)$, whose energy density and pressure read
\begin{equation}
\rho_\phi=\frac{\phi'^2}{2\,a^2}+V(\phi)\qquad , \qquad p_\phi=\frac{\phi'^2}{2\,a^2}-V(\phi).
\end{equation}
In the previous equations the prime stands for derivative with respect to the conformal time.  This scalar field can be mapped to the perfect fluid with equations of state (\ref{p}) or (\ref{p2}); i.e. $\rho_\phi=\rho$ and $p_\phi=p$. Through this mapping we can write down $\phi$ and $V$ as a function of the scale factor (see Fig.~\ref{phixa}):
\begin{eqnarray}
\phi(a)&=&\frac{1}{q\kappa}\left\{4\tanh^{-1}\sqrt{1+\frac{q}{4(1+\alpha)}\frac{1}{1+x}}-2\sqrt{\frac{1+\beta}{1+\alpha}}\right.\nonumber\\
&\,&\left.\coth^{-1}\left[\sqrt{\frac{4(1+\alpha)}{1+\beta}\left(1+\frac{q}{4(1+\alpha)}\frac{1}{1+x}\right)}\,\,\right]\right\},\nonumber\\
\label{phia}\\
V(a)&=&A^{1/(1+\alpha)}\left(\frac{A}{B}\right)^{-(1+\beta)/[q(1+\alpha)]} x^{-(1+\beta)/[q(1+\alpha)]} \nonumber\\
&\,& (1+x)^{1/(1+\alpha)}\left[\frac13-\frac{q}{6(1+\alpha)}\frac{1}{1+x}\right],\label{Va}
\end{eqnarray}
where $x=(B/A)a^q$ and $q=1+\beta-4(1+\alpha)$. The scalar field starts with a negative value and it rolls down the potential as the universe inflates (cf. Fig.~\ref{V1}).

\begin{figure}[b]
\includegraphics[width=8cm]{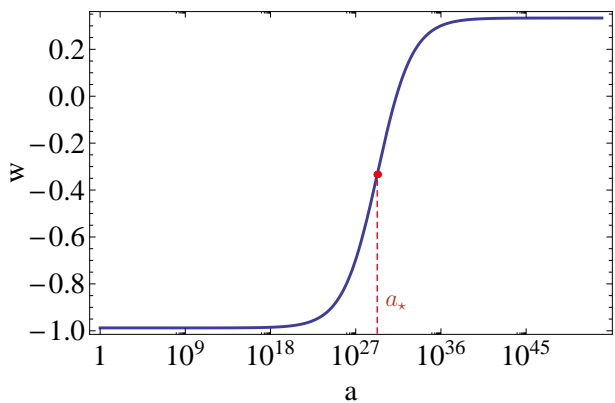}
\caption{The equation of state $w$ for the modified GCG model with equation of state (\ref{p}) or (\ref{p2}). The dashed red line indicates the end of inflation. We set $\alpha=-1.06$ as an example here.}
\label{w}
\end{figure}

We can also consider the behavior of the equation of state parameter $w$, $w=p_\phi/\rho_\phi$, for the scalar field as a function of the scale factor (see Fig.~\ref{w}). At early times, the kinetic energy is negligible with respect to the potential, and therefore $w\rightarrow -1$. Afterwards, the parameter $w$ increase gradually with time. The universe stops inflating when $w$ reaches the value $-1/3$ and becomes radiation dominant when $w$ finally reaches $1/3$.

At very early times where the scale factor is very small, it can be shown that the inflationary expansion follows a power law. Indeed, the scalar field and the potential can be approximated by
\begin{eqnarray}
\phi(a)&\simeq&\phi_0+\frac{1}{\kappa}\sqrt{\frac{1+\beta}{1+\alpha}}\ln(a),\\
V(a)&\simeq&\left[1-\frac{1+\beta}{6(1+\alpha)}\right] A^{1/(1+\alpha)}a^{-(1+\beta)/(1+\alpha)},\\
\phi_0&=&\frac{1}{q\kappa}\left[\sqrt{\frac{1+\beta}{1+\alpha}}\ln\left(\frac{qB}{4|1+\alpha|A}\right)\right.\nonumber\\
&\,& \left.+4\tanh^{-1}\sqrt{\frac{1+\beta}{4(1+\alpha)}}\,\,\right],
\end{eqnarray}
and therefore
\begin{eqnarray}
V(\phi)\simeq\left[1-\frac{1+\beta}{6(1+\alpha)}\right]A^{1/(1+\alpha)}\exp\left[-\kappa\sqrt{\frac{1+\beta}{1+\alpha}}\left(\phi-\phi_0\right)\right].\nonumber\\
\label{Vapowerlow}
\end{eqnarray}

At much later times where the scale factor becomes large, the universe is radiation dominant. It can be easily proved that for this period:
\begin{eqnarray}
\phi(a)&\simeq& \frac{2}{\kappa}\ln(a)+\phi_1, \\
V(a)&\simeq& \frac13 B^{1/(1+\alpha)}a^{-4},\\
\phi_1&=&-\frac{2}{\kappa q} \left[\ln\left(\frac{Aq}{16B|1+\alpha|}\right)\right.\nonumber \\
&\,&\left.+\sqrt{\frac{1+\beta}{1+\alpha}}\coth^{-1}\sqrt{\frac{4(1+\alpha)}{1+\beta}}\,\right].
\end{eqnarray}
Consequently, we obtain
\begin{equation}
V(\phi)\simeq\frac13 B^{1/(1+\alpha)}\exp\left[-2\kappa(\phi-\phi_1)\right].
\label{powerlawpotential}
\end{equation}

Since we know the form of the potential $V(\phi)$ for this modified GCG model, we are able to find the evolution of the slow-roll parameters $\epsilon(a)$ and $\eta(a)$, where
\begin{equation}
\epsilon=\frac1{2\kappa^2}\left(\frac{1}{V}\frac{dV}{d\phi}\right)^2,\quad \eta=\frac1{\kappa^2}\frac{1}{V}\frac{d^2V}
{d\phi^2}.
\end{equation}

\begin{figure}[b]
\includegraphics[width=7.5cm]{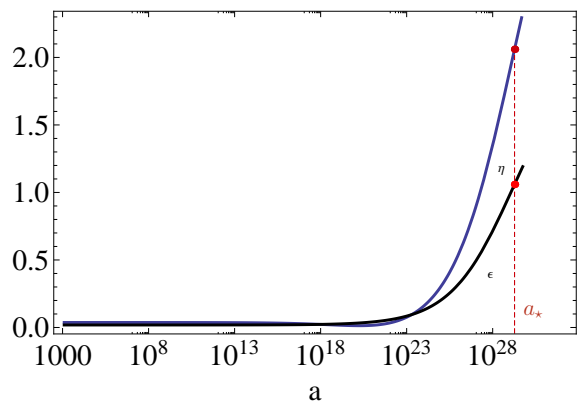}
\caption{The black line corresponds to $\epsilon$ and the blue one to $\eta$. The dashed red line locates the time when inflation ended. We also notice that the slow-roll conditions are no longer satisfied when the field is close to the end of inflation ($a=a_*$). We set $\alpha=-1.06$ as an example here.}
\label{slowroll}
\end{figure}

The slow-roll approximation is valid when the conditions $\epsilon\ll1,\,\eta\ll1$ are satisfied, which means the potential energy dominates over the kinetic term during the inflation era. In Fig.~\ref{slowroll}, we show the behavior of these functions in terms of the scale factor $a$.

\section{Primordial power spectrum}

Inflation not only solve some of the shortcomings present in the big bang theory but also generates density perturbations that seeds the structure of the present universe. Those density perturbations have been constrained through observations of the cosmic microwave background (CMB). In this section, we will constrain the model introduced in the previous section by using the measurements of  WMAP7 \cite{Komatsu:2010fb}
 for the  power spectrum of the comoving curvature perturbations, $P_s=2.45\times 10^{-9}$, and its index, $n_s=0.963$, where
\begin{equation}
n_s-1\equiv\frac{d\ln P_{s}(k)}{d\ln k}.
\label{defns}
\end{equation}
These measurements correspond to a pivot scale $k_0=0.002\, {\mathrm{Mpc^{-1}}}$ \cite{Komatsu:2010fb}.

Once the parameters of the model have been constrained we will obtain the full spectrum of the scalar perturbations.

\subsection{Imposing Observational Constraints}

At very early times, the model introduced in Sect.~II induces a power law expansion as shown clearly in Eq.~(\ref{powerlaw}) (see also Eq.~(\ref{Vapowerlow}) and Fig.~\ref{V1}). On the other hand, the slow-roll approximation is valid for an inflationary power law expansion as the one we are considering (see for example Figs.~\ref{w} and \ref{slowroll}). In this regime, the power spectrum (for the comoving curvature perturbations) and the spectral index can be expressed as (see for example \cite{Langlois:2010xc,Bassett:2005xm,Lidsey:1995np})
\begin{eqnarray}
P_s&\simeq&\left(\frac{H^2}{2\pi \dot{\phi}}\right)^2,\label{pssr}\\
n_s&\simeq& 1-6\epsilon+2\eta,\label{nssr}
\end{eqnarray}
at the horizon exit; i.e. at $k=aH$. Throughout the paper a dot stands for a derivative with respect to the cosmic time. For a power law expansion, i.e., $a\propto \tau^l$ where $\tau$ is the conformal time, the spectral index reduces to $n_s=-(1+2/l)$. For the modified Chaplygin gas we are analyzing $l^{-1}=(1+\beta)/2(1+\alpha)-1$.

With the previous inputs, we can proceed as follows: (i) the parameter $B$ is fixed by the current amount of radiation in the universe as stated in Eq.~(\ref{Bdef}), (ii) for a given parameter $\alpha$, the parameter $\beta$ is fully determined by the measurement of $n_s$ and (iii) the parameter $A$ is fixed such that $P_s=2.45\times 10^{-9}$ at the pivot scale $k_0=0.002\, {\mathrm{Mpc^{-1}}}$. Before moving forwards, we would like to highlight that the mode $k_0=0.002\, {\mathrm{Mpc^{-1}}}$ exits the horizon well inside the inflationary era where both the power law expansion and the slow-roll approximation are valid for the model (\ref{p2}). We found out that the inflationary scale, $V_0=A^{1/(1+\alpha)}(A/B)^{-(1+\beta)/q(1+\alpha)}\sim 1.2\times 10^{16}\,\, \rm{GeV}$, is almost of constant.

The tensor power spectrum to the scalar power spectrum ratio in the slow-roll approximation can be expressed through the slow-roll parameter $\epsilon$ \cite{Langlois:2010xc,Bassett:2005xm,Lidsey:1995np}:
\begin{equation}
r\equiv\frac{P_t(k)}{P_s(k)}\approx16\epsilon .
\end{equation}
For power-law inflation in the slow-roll approximation, the spectral index is $n_s=-(1+2/l)$, as we already mentioned, while $\epsilon=1+1/l$. Because this GCG model behaves like a slow-roll power-law inflation when the pivot $k_0$ exits the horizon, we can use these approximations to evaluate $r$ at that time. Using once more WMAP7 data ($n_s\approx0.963$ at the scale $k_0=0.002\textrm{Mpc}^{-1}$) we can find $l$. We obtain $r=0.296$ at the pivot scale $k_0$, which is in agreement with the WMAP7 constraints. We obtain the full spectrum of the gravitational waves on section V.

At later times close to the end of inflation or when larger modes $k$ exit the horizon, the slow-roll conditions are not fulfilled. This is a simple consequence of the behavior of the GCG model that deviates from a power-law inflation at that time. Thus we cannot use  the results  in Eqs.(\ref{pssr}) and (\ref{nssr}) when inflation approaches $a_*$ where it halts. In this regime, we will rather calculate the spectrum numerically.

\subsection{Primordial Power Spectrum: Methods and Results}

The primordial density perturbation can be directly mapped to the comoving curvature perturbation, the last remains constant on large scale after exiting the Hubble horizon, as long as the perturbations are adiabatic. This is precisely the case, as we have a unique degree of freedom corresponding to the generalised Chaplygin gas.

The comoving curvature perturbation, $s$, is determined by the fluctuations of the generalised Chaplygin gas. The corresponding power spectrum for the field $\nu_k$ is \cite{Langlois:2010xc,Bassett:2005xm}
\begin{equation}
2\pi^2k^{-3}P_s(k)=\frac{|\nu_k|^2}{z^2}\label{powerspectrum},
\end{equation}
where $z=\frac{a\dot{\phi}}{H}$. The field $\nu_k$ satisfies in the Fourier space  the equation \cite{Langlois:2010xc,Bassett:2005xm}
\begin{equation}
\nu_k^{\prime\prime}+\left(k^2-\frac{z^{\prime\prime}}{z}\right)\nu_k=0\label{seof}.
\end{equation}
If the slow-roll conditions are satisfied, the variation of the field $\phi$ and the Hubble parameter $H$ is much slower than that of the scale factor $a$. Therefore, the following approximation can be used: $z^{\prime\prime}/z\approx a^{\prime\prime}/a$, at the lowest order of the slow-roll approximation. Consequently, the solutions of  (\ref{seof}) read in this case \cite{Langlois:2010xc}
\begin{equation}
\nu_k\approx\sqrt{\frac{1}{2k}}\,e^{-ik\tau}\left(1-\frac{i}{k\tau}\right) \label{nusol},
\end{equation}
where the Bunch-Davies vacuum has been imposed for modes well inside the horizon ($aH\ll k$). We remind that the parameter $\tau$ stands for the conformal time. Now, if we consider the super-horizon scale $(|k\tau|\ll1)$ and substitute the approximation $a\approx-1/(H\tau)$ into Eq.~(\ref{nusol}), the solution becomes
\begin{equation}
\nu_k\approx i\sqrt{\frac1{2k}}\frac{aH}{k}.
\end{equation}
The corresponding power spectrum in this slow-roll approximation is given in Eq.~(\ref{pssr}). These results remain valid on the next order of the slow-roll as shown for example in \cite{Bassett:2005xm,Lidsey:1995np} and therefore can be applied for an inflationary power law expansion as it is the case in the model we are analyzing at very early time.

In order to obtain the full power spectrum $P_s(k)$, we use numerical methods instead of applying Eq.~(\ref{pssr}) for the reasons stated in the previous subsection. Therefore, we first solve the differential equation (\ref{seof}), and find the solution $\nu_k(a_{\rm{cross}})$ calculated at the time when the mode $k$ exits the horizon; i.e. $k=a_{\rm{cross}}H$. Then through Eq.~(\ref{powerspectrum}) we obtain the power spectrum $P_s(k)|_{k=aH}$.

We proceed as follows. We separate Eq.~(\ref{seof}) into two first order differential equations:
\begin{equation}
\left\{\begin{array}{ll}
X^{\prime}=Y ,\\
Y^{\prime}=-\left(k^2-\frac{z^{\prime\prime}}{z}\right)X,
\end{array}\right.
\end{equation}
where we set $X=\nu_k$. In order to solve the previous set of differential equations, we take the following actions{\color{blue}:}
\begin{itemize}
\item  It is easier to make a change of variable from the conformal time to the scale factor. Notice that $z=a\dot{\phi}/H$ can be fully determined in terms of the scale factor
as $\dot\phi=aH d\phi/da$ and the scalar field $\phi$ is fully determined as a function of the scale factor (see Eq.~(\ref{phia})). \item In addition, we have simply to use the set of values for the parameters of our model that have been deduced by imposing observational constraints on the model as discussed in the previous subsection.
\item Last but not least, we need to impose a set of boundary conditions. When the wavelength of a given mode $k$ is much smaller than the Hubble radius\footnote{This condition is fulfilled by any mode (on the past) thanks to the inflationary mechanism.} $k\gg aH$, the effect of curvature can be neglected. Therefore, the result reduces to that of a  flat Minkowski spacetime (when $k\gg aH$). So, the initial condition is
\begin{equation}
\nu_k\rightarrow\sqrt{\frac1{2k}}\, e^{-ik\tau} \quad\mbox{for}\,\, |k\tau|\gg1 \label{Minkowski},
\end{equation}
which corresponds to the Minkowski vacuum at very early time. We change the variable from the conformal time $\tau$ to the scale factor $a$  through the relation\footnote{Notice at this regard that the expansion of the universe for modes with a wavelength much smaller than the Hubble radius is well approximated by a power law.}
\begin{equation}
a(\tau)=\left\{\left[\frac{1+\beta}{2(1+\alpha)}-1\right]\sqrt{\frac{8\pi G}{3}}A^{1/[2(1+\alpha)]}\tau\right\}^{1/\left[\frac{1+\beta}{2(1+\alpha)}-1\right]}. %\quad(\mbox{power-law inflation}),
\label{atau}
\end{equation}
\end{itemize}

\begin{figure}[t]
\includegraphics[width=9cm]{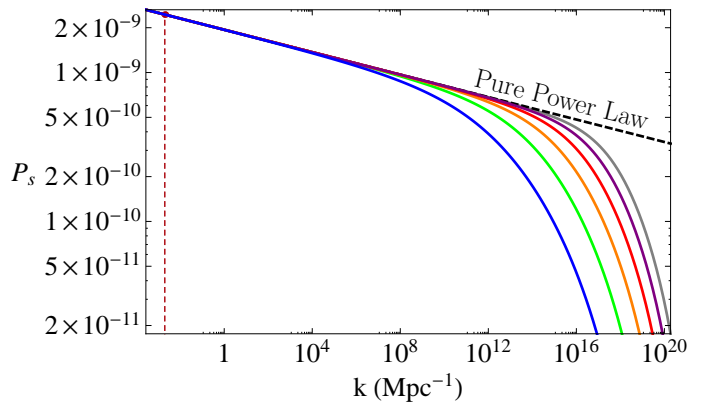}
\caption{Primordial power spectrum $P_s(k)|_{k=aH}$ against $k$ for six different values of $\alpha$. The dashed black line is the pure power law inflation, and the vertical dashed red line locates the pivot $k_0=0.002\textrm{Mpc}^{-1}$. We can see that all these lines merge when small $k$; i.e. large scale, exits the horizon. The grey, violet, red, orange, green and blue curve correspond respectively to $\alpha=-1.1,-1.09,-1.08,-1.07,-1.06,-1.05.$}
\label{Ps}
\end{figure}

\begin{figure}[b]
\includegraphics[width=8.2cm]{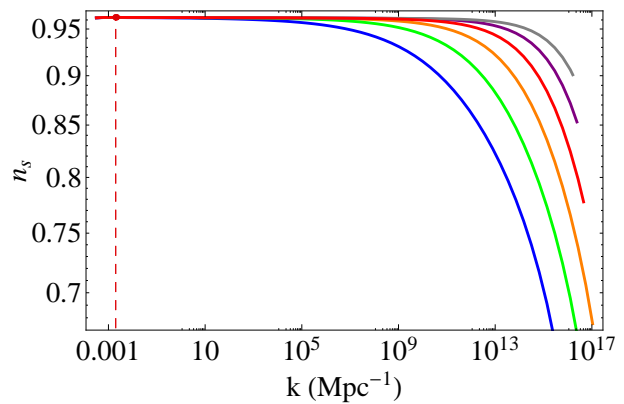}
\caption{The spectral index $n_s$ against $k$. The red dashed line locates the pivot scale $k_0=0.002\textrm{Mpc}^{-1}$. The spectral index $n_s$ is approximately a constant at early time. The grey, violet, red, orange, green and blue curve correspond respectively to $\alpha=-1.1,-1.09,-1.08,-1.07,-1.06,-1.05.$}
\label{ns}
\end{figure}

Following a similar procedure, we can see if the spectral index, $n_s$ (defined in Eq.~(\ref{defns})), depends on the scale. Our results are shown in Figs.~\ref{Ps} and \ref{ns}.
%%%%%
%%%%%

For lower modes, i.e., those that entered the horizon very recently, the power spectrum is independent of the parameter\footnote{The other parameters of the model are fixed for a given $\alpha$ as explained in the previous subsection.} $\alpha$.  It is only for modes satisfying $10^5\mathrm{Mpc}^{-1}\leq k$ that we can start to see some dependence of $P_s$ on $\alpha$. As should be expected, the power spectrum has a constant slope; i.e., a constant spectral index, for lower $k$ when the modified Chaplygin gas induces a power-law expansion. Indeed, the results for power-law matches very well those of our model for these modes (cf. Figs.~\ref{Ps} and \ref{ns}). However, the spectral index of the primordial power spectrum is not scale independent as Fig.~\ref{ns} clearly shows. The deviation of $n_s$ from the scale independence starts when the slow-roll condition ceases to be valid and therefore approximately a  bite before the end of inflation.

Before concluding this section, we recall that the current CMB observations cover roughly the range of scale $k\approx0.0002\textrm{Mpc}^{-1}$ to $k\approx0.2\textrm{Mpc}^{-1}$ \cite{Zhao:2010ic}. In this observable region, it is hard to separate a modified GCG model for different values of $\alpha$ or even from a power-law inflation. Could currently running observations or future missions help us in this regard? For example, it is expected that the maximum scale $\ell$ in CMB observations from the Planck mission would reach $\ell_{max}\approx2000$. This scale $\ell_{max}$ could be translated into a $k_{max}$ scale. Even though the relation between $\ell$ and $k$ is not a one-to-one relationship, we can still find the main contribution for each mode: $k\left(\tau_0-\tau_*\right)\sim\ell$ \cite{1}, where $\tau_0-\tau_*$ is the elapsed conformal time since the last scattering surface until the present. Here we use the concordance $\Lambda$CDM model to calculate $\left(\tau_0-\tau_*\right)=\int^{a_0}_{a_*}da/a^2H$ numerically. We find that the maximum scale corresponding to $\ell_{max}\approx2000$ is $k_{max}\approx0.138\textrm{Mpc}^{-1}$. Consequently and unfortunately, the answer to the question raised in this paragraph is negative, as is clear from Figs.~\ref{Ps} and \ref{ns}. We will next show that by looking at the high frequency spectrum of the gravitational waves, it is possible to separate a modified GCG model for different values of $\alpha$. Those frequencies might be within the reach of future gravitational-wave detectors such as BBO and DECIGO \cite{Lidsey97}.

\section{Gravitational Wave Spectrum}

The early inflationary era not only leaves imprints on the scalar cosmological perturbations but also creates a fossil of gravitational waves. In this section, we will analyze the possible imprints in the power spectrum of the stochastic background of gravitational waves, for the model presented in section II. In this case the background evolution of the universe, till nowadays, is described by the matter content in Eq.~(\ref{energydensities}). This analysis is quite important as it can shed some light on the inflationary scenario behind the early accelerating phase of the universe and its transition to the subsequent radiation era.

\subsection{Method}

In order to obtain the spectrum of the gravitational waves (GWs), we will use the method of Bogoliubov coefficients. To our knowledge, this method was first developed in \cite{Parker:1969au,Starobinsky,allen}  and later applied in \cite{Moorhouse:1994nc,Mendes:1994ai,Sa:2008yq,Sa:2007pc}. Bogoliubov coefficients describe how the vacuum changes as the universe expands. In particular, one of these coefficients, which we will denote $\beta_k$, gives the number of gravitons created. Furthermore, it can be shown that the dimensionless relative logarithmic energy spectrum of the gravitational waves, $\Omega_{\mathrm{GW}}$, at the present time reads \cite{Sa:2008yq,Sa:2007pc}:
\begin{equation}
\Omega_{GW}(\omega,\tau_0)\equiv\frac{1}{\rho_c(\tau_0)}\frac{d\rho_{GW}}{d\ln\omega}(\tau_0)=\frac{\hbar\kappa^2}{3\pi^2 c^5 H^2(\tau_0)}\omega^4\beta_k^2(\tau_0).
\label{spectrum}\end{equation}
The parameter $\rho_{\rm GW}$ is the energy density of GWs and  $\omega$  the respective
 angular frequency; $\rho _{\rm c}$ and $H$ are the critical density of the
universe and Hubble parameter, respectively, evaluated at the present time.

In summary, the present value of the Bogoliubov coefficient, $\beta_k$, determines the power spectrum of the gravitational waves. This parameter can be determined in terms of two continuous functions $X,Y$ such that $|\beta_k|^2=(X-Y)^2/4$, where $X,Y$ fulfil
\begin{equation}
\left\{\begin{array}{ll}X^\prime=-ikY \\  Y^\prime=-\frac{i}{k}\left(k^2-\frac{a^{\prime\prime}}{a}\right)X\end{array}\right..\label{ODE}
\end{equation}
The coefficient $\beta_k$ gives the number of gravitons, $N_k$, for each mode $k$ created during the evolution of the universe, where $N_k(\tau)=|\beta_k(\tau)|^2$.

\begin{figure}[t]
\includegraphics[width=8.3cm]{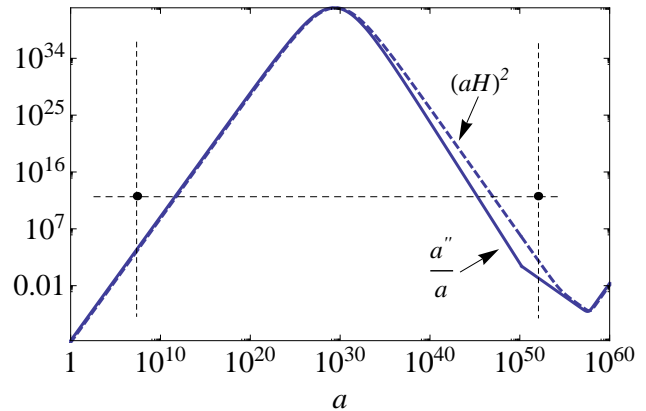}
\caption{The integration method we use for a given mode $k$. The dashed line corresponds to $(aH)^2$, and the solid line corresponds to $a^{\prime\prime}/a$ for a modified GCG model.}
\label{integration}
\end{figure}

To calculate the GWs spectrum (\ref{spectrum}), we have to solve the differential equations (\ref{ODE}) numerically and use appropriate initial conditions for $X(\tau_i)$ and $Y(\tau_i)$. We proceed as follows:
\begin{itemize}
\item The expression for $a^{\prime\prime}/a$ can be obtained from the Friedmann equation and the conservation law:
\begin{equation}
\frac{a^{\prime\prime}}{a}=\frac{\kappa^2}6a^2\left(\rho-3p\right),
\end{equation}
and therefore can be written in terms of the scale factor, $a$, in this modified GCG model as follows,
\begin{equation}\label{potentialgw}
\frac{a^{\prime\prime}}{a}=\left\{
\begin{array}{ll}
\frac{\kappa^2}{6}a^2\left[4-\left(\frac{1+\beta}{1+\alpha}\right)\right]\frac{A}{a^{1+\beta}}\times&\\
\left(\frac{A}{a^{1+\beta}}+\frac{B}{a^{4(1+\alpha)}}\right)^{-\alpha/(1+\alpha)}\,,     &\mbox{early time}\\
\frac{\kappa^2}{6}a^2\left[\rho_{m0}\left(\frac{a_0}{a}\right)^3+4\rho_\Lambda\right]\,, &\mbox{late time}
\end{array}\right.
\end{equation}
\item At very early time, the model we are analyzing behaves like a pure power-law inflation. In that case, the set of differential equations (\ref{ODE}) have an analytical solutions \cite{Sa:2008yq}, which we will use as initial conditions for $X(\tau)$ and $Y(\tau)$ in our numerical integration\footnote{Notice that the conformal time is negative.}:
\begin{align}
&X(\tau_i)=\sqrt{\frac{-k\tau_i\pi}{2}}H_{\frac12-p}^{(1)}(-k\tau_i),\label{BCGW1}\\
&Y(\tau_i)=-i\sqrt{\frac{-k\tau_i\pi}{2}}\left[H_{-\frac12-l}^{(1)}(-k\tau_i)+\frac{l}{-k\tau_i}H_{\frac12-l}^{(1)}(-k\tau_i)\right].
\label{BCGW2}
\end{align}
Before proceeding, let us highlight that the general solution of Eq.~(\ref{ODE}) for a power-law inflation includes the first and the second kind of Hankel functions \cite{Abramowitz}. But here we adopt just Hankel function of the first kind because the solution should reduce to the Minkowski spacetime solution (\ref{Minkowski}) at very early time when $aH\ll k$ \cite{Lidsey:1995np}.

\item Similarly to the scalar perturbations, it is simpler to solve the set of equations (\ref{ODE}) in terms of the scale factor. With this in mind and in order to apply the boundary conditions (\ref{BCGW1}) and (\ref{BCGW2}), we use the relation between the conformal time and the scale factor stated in Eq.~(\ref{atau}).

\item An important issue is the range of integration of Eqs.~(\ref{ODE}). In principle, the integration must be done from a very early time when the mode is well inside the horizon on the inflationary era to the present time. However, there is a way of reducing the computing time without affecting in practice the results: for a given mode $k$, we start integrating from 0.01$a_{\mathrm{cross}_1}$ until 100$a_{\mathrm{cross}_2}$, where $a_{\mathrm{cross}_1}$ and $a_{\mathrm{cross}_2}$ correspond to the scale factor where the mode $k$ exits the horizon and reenters the horizon, respectively. Our numerical integration will be carried out for modes $k$ ranging from the maximum of the potential $a''/a$ until the minimum one. The last one corresponding roughly to the modes that are reentering currently  the horizon. The methods is schematically shown in Fig.~\ref {integration}.

\item As can be noticed from Fig.~\ref {integration}, there is an intersection point in the potential $a''/a$ which takes place roughly at $10^{50}$. Given that we have set the current scale factor to $a_0=10^{58}$, the intersection point corresponds roughly to the time of Big Bang Nucleosynthesis (BBN). On our model, that point corresponds to the transition between the modified generalized Chaplygin gas and the  $\Lambda$CDM model. For modes that enters the horizon after the BBN, we will have to integrate Eqs.~(\ref{ODE}) twice as we will be dealing with two different potentials as shown in Eq.~(\ref{potentialgw}). For the second integration, we use the final values obtained in the first integration as the initial values for the functions $X$ and $Y$.

\item The integration will be performed assuming: $\Omega_{r0}=8\times10^{-5}$, $\Omega_{m0}=0.24$, $a_0=10^{58}$  and $H_0=71.0$ km/s/Mpc. In addition, the parameters of the model are fixed as explained in section II.

\end{itemize}

\subsection{Results}

In this subsection we present our results of the power spectrum of the gravitational waves for the model presented in section II.

\begin{figure}[t]
\centering
\includegraphics[width=8cm]{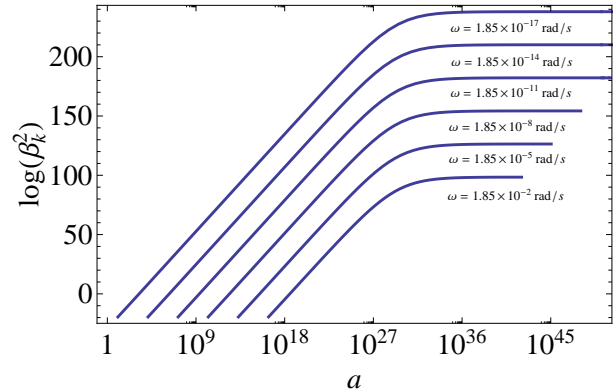}
\caption{Evolution of the graviton numbers, $|\beta_k|^2=N_k$. The integration has been carried out for six different values of frequencies. We can see that for each mode $k$, the gravitons have been created mostly during the inflationary era. For this plot we fixed $\alpha=-1.06$.}
\label{betasqure}
\end{figure}

As we explained before, the Bogoliubov coefficient, $\beta_k$, gives the number of graviton created in each period, $|\beta_k|^2=N_k(\tau)$, and therefore we can calculate the time evolution of $\beta_k^2$ during the expansion of the Universe in this GCG model. In Fig.~\ref{betasqure} we present an example of our results where we set $\alpha=-1.06$. The plot corresponds to six different wave numbers $k$: $k=1.9\times10^{-3} \textrm{Mpc}^{-1}$, $k=1.9 \textrm{Mpc}^{-1}$, $k=1.9\times10^3 \textrm{Mpc}^{-1}$, $k=1.9\times10^6 \textrm{Mpc}^{-1}$, $k=1.9\times10^{9} \textrm{Mpc}^{-1}$, and $k=1.9\times10^{12} \textrm{Mpc}^{-1}$, or different frequencies $\omega=k/a_0$.

\begin{figure}[t]
\centering
\includegraphics[width=9cm]{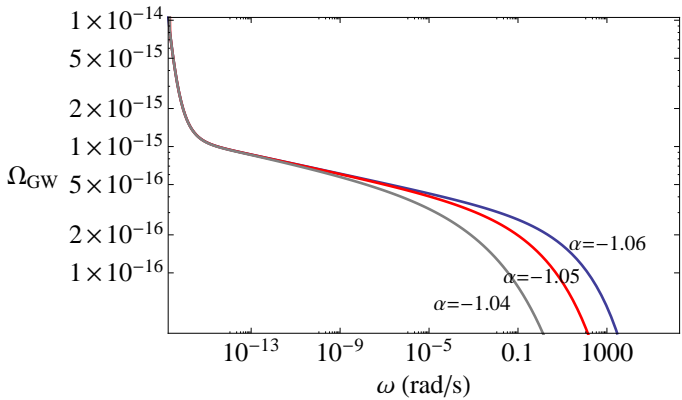}
\centering
\caption{The gravitational wave spectrum $\Omega_{GW}$ against the frequency $\omega$ for different value of $\alpha$ in this GCG model: the blue line refers to $\alpha=-1.06$, the red one refers to $\alpha=-1.05$, and the grey one to $\alpha=-1.04$.}
\label{Gwspectrum}
\end{figure}

Similarly we can obtain the spectrum of the gravitational waves as shown in Fig.~\ref{Gwspectrum}. This figure is quite enlightening. It shows that for large frequencies the larger $|\alpha|$ is, the larger is the dimensionless logarithmic energy spectrum of the GWs. On the other hand, the larger $|\alpha|$ is, the more the spectrum is shifted towards larger frequencies. This is a simple consequence of the fact that a larger $|\alpha|$ implies a larger maximum of the potential $a''/a$. In summary, an increase in $|\alpha|$ implies two things: an upward shift and a rightward shift of the spectrum. It is also worthy of notice that the plateau of the spectrum merges at middle/low frequencies for different values of $\alpha$. This is not surprising as the energy scale of inflation, $V_0$, is almost fixed and independent of $\alpha$ in our model, where $V_0\sim 1.2\times 10^{16} \rm{GeV}$ (cf. the previous section) and it is precisely the value of $V_0$  that shifts vertically such a plateau \cite{Sa:2008yq}. The parameter $V_0$ is fixed mainly by the measurements of $P_s$ and $n_s$. The fact that $n_s$ is much more constrained by the latest WMAP7 data than the previous one, WMAP5, explains the changes in the plateau of our Fig.~\ref{Gwspectrum} and Fig.~4 of Ref.~\cite{Sa:2008yq}, where on the last mentioned figure different values of $n_s$, or equivalently different values of the power law universal expansion $l$ (in terms of the conformal time) are assumed. Therefore, the spectrum of GWs at very low frequencies is insensitive to the value of $\alpha$, because  for larger frequencies the spectrum already merged for different $\alpha$'s whereas for smaller frequencies the universe is described by the concordance model $\Lambda$CDM (at time the modes reenters the horizon), which is independent of the parameter $\alpha$.

There are some observational constraints on the upper limit of the energy spectrum $\Omega_{GW}$ \cite{Sa:2007pc}:
\begin{itemize}
\item
Constraint from CMB: \\$h_0^2\Omega_{GW}(\omega_{hor},\tau_0)\leq7\times10^{-11}, \\ \textrm{for} \omega_{hor}=2\times10^{-17}h_0\,\,rad/s$.
\item
Constraint from timing observations of millisecond pulsars:  \\$h_0^2\Omega_{GW}(\omega_{pul},\tau_0)<2.0\times10^{-8},\\ \textrm{for}\quad \omega_{pul}=2.5\times10^{-8}\,\,rad/s$.
\item
Constraint from doppler tracking of the Cassini spacecraft: \\ $h_0^2\Omega_{GW}(\omega_{Cas},\tau_0)<0.014, \\ \textrm{for}\quad \omega_{Cas}=7.5\times10^{-6}\,\,rad/s$.
\item
Constraint from LIGO:  \\ $h_0^2\Omega_{GW}(\omega,\tau_0)<3.4\times10^{-5}$, \\ for a few hundred $rad/s$.
\item
Constraint from BBN: \\ $h_0^2\int_{\omega_n}^\infty\Omega_{GW}(\omega,\tau_0)d\omega/\omega<5.6\times10^{-6}$, \\ where
$\omega_n\approx10^{-9}\,\,rad/s$,
\end{itemize}
\noindent where $h_0=0.71$. All these constraints are fulfilled by the power spectrum shown in Fig.~\ref{Gwspectrum}.

This model is a clear example of how the detection of GWs can be extremely helpful to distinguish among several inflation models and a complementary tool to the study of primordial spectrum of the scalar perturbations. Indeed, while our model is indistinguishable for any value of $\alpha$ at low $k$ and $\omega$, on the spectrum of scalar and tensorial perturbations, respectively, it should be distinguishable at high $k$ and $\omega$. Most importantly, the high frequency regime is within the reach of future gravitational wave detectors such as BBO and DECIGO \cite{Lidsey97}.

\section{conclusions}

We present a model that attempts to fuse the inflationary era and the subsequent radiation dominated era under a unified framework so as to provide a smooth transition between the two. The model is based on a modification of the generalized Chaplygin gas. More precisely, it interpolates between a ``quintessence-like'', or power law, expansion and a subsequent radiation dominated universe. Such a gas fulfills an equation of state (\ref{p}) or (\ref{p2}) which has been previously used in the literature as a means to explain a possible interplay between the dark sectors of the Universe \cite{Chimento:2009sh}. Our attempt is therefore in a different context. We have shown how such a model has an underlying scalar field description, where the scalar field starts rolling down a potential hill where the extreme slow-roll approximation is valid until a bite before the end of inflation.

We have obtained the full power spectrum of the scalar perturbations and constrained the model using the latest WMAP7 data. From our analysis it turns out that the model is indistinguishable from a power-law expansion at low wave numbers $k$ (cf.~Figs.~\ref{Ps} and \ref{ns}) even if we vary the only degree of freedom present on the model, the parameter $\alpha$. Notice in this regard that the rest of the parameters of the model are fixed by observations as explained on Sect.~IV.

We have completed our analysis by obtaining the spectrum of the gravitational waves (cf.~Fig.~\ref{Gwspectrum}) where we have used the method of Bogoliubov coefficients  \cite{Parker:1969au,Starobinsky,allen,Moorhouse:1994nc,Mendes:1994ai,Sa:2008yq,Sa:2007pc}. It turns out that our model is within the reach of future gravitational-wave detectors like BBO and DECIGO \cite{Lidsey97}. Most importantly, for those frequencies it appears possible to distinguish this modified GCG model among different values of $\alpha$, while this seems unattainable using the power spectrum of the scalar perturbation with the present or near future observations.

Last but not least, we have shown as well how the scalar perturbation spectrum of the present model is not as red as the one presented in \cite{BouhmadiLopez:2009hv} for an alternative GCG model for the early universe. In comparison, the model presented here is more consistent with observations.

\acknowledgments

The authors are grateful to Alfredo B. Henriques for a careful reading of the manuscript.
M.B.L. is  supported by the Portuguese Agency Funda\c{c}\~{a}o para a Ci\^{e}ncia e Tecnologia through the fellowship SFRH/BPD/26542/2006. She also wishes to acknowledge the hospitality of LeCosPA Center at the National Taiwan University during the completion of part of this work. P.C. and Y.W.L. are supported by Taiwan National Science Council under Project No. NSC 97-2112-M-002-026-MY3 and by Taiwan's National Center for Theoretical Sciences (NCTS). P.C. is in addition supported by US Department of Energy under Contract No. DE-AC03-76SF00515.

\end{document}